\documentclass[aps,prd,twocolumn,preprintnumbers,superscriptaddress,dblfloatfix,nofootinbib]{revtex4-1}
\usepackage[utf8]{inputenc}
\usepackage{lmodern}
\usepackage{amsmath,amssymb,mhchem}
\usepackage{graphicx}
\usepackage{url}
\usepackage{color}
\usepackage{enumitem}
\usepackage{subfigure}
\usepackage[dvipsnames]{xcolor}
\usepackage[colorlinks=true,breaklinks=true]{hyperref}
\hypersetup{allcolors=[rgb]{0.0 0.0 0.6},linkcolor=[rgb]{0.75 0.05 0.05}}
\usepackage{bm}
\usepackage{epsfig}
\usepackage{amsmath}
\usepackage{amssymb}
\usepackage{slashed}
\usepackage{color}
\usepackage{accents}
\usepackage[dvipsnames]{xcolor}
\usepackage[colorlinks=true,breaklinks=true]{hyperref}
\hypersetup{allcolors=[rgb]{0.0 0.0 0.6},linkcolor=[rgb]{0.75 0.05 0.05}}
\usepackage[export]{adjustbox}

\begin{document}

\preprint{KEK-QUP-2023-0021, KEK-TH-2553, KEK-Cosmo-0323, IPMU23-0031}

\title{Primordial Black Hole Neutrinogenesis of Sterile Neutrino Dark Matter
}

\author{Muping Chen}
\email{mpchen@physics.ucla.edu}
\affiliation{Department of Physics and Astronomy, UCLA,\\
475 Portola Plaza, Los Angeles, CA 90095, USA}

\author{Graciela B. Gelmini}
\email{gelmini@physics.ucla.edu}
\affiliation{Department of Physics and Astronomy, UCLA,\\
475 Portola Plaza, Los Angeles, CA 90095, USA}

\author{Philip Lu}
\email{philiplu11@gmail.com}
\affiliation{Center for Theoretical Physics, Department of Physics and Astronomy, Seoul National University, Seoul 08826, Korea}
\affiliation{International Center for Quantum-field Measurement Systems for Studies of the Universe and Particles (QUP, WPI),
High Energy Accelerator Research Organization (KEK), Oho 1-1, Tsukuba, Ibaraki 305-0801, Japan}

\author{Volodymyr Takhistov}
\email{vtakhist@post.kek.jp}
\affiliation{International Center for Quantum-field Measurement Systems for Studies of the Universe and Particles (QUP, WPI),
High Energy Accelerator Research Organization (KEK), Oho 1-1, Tsukuba, Ibaraki 305-0801, Japan}
\affiliation{Theory Center, Institute of Particle and Nuclear Studies (IPNS), High Energy Accelerator Research Organization (KEK), Tsukuba 305-0801, Japan
}
\affiliation{Graduate University for Advanced Studies (SOKENDAI), \\
1-1 Oho, Tsukuba, Ibaraki 305-0801, Japan}
\affiliation{Kavli Institute for the Physics and Mathematics of the Universe (WPI), UTIAS, \\The University of Tokyo, Kashiwa, Chiba 277-8583, Japan}

\date{\today}

\begin{abstract}
Sterile neutrinos are well-motivated and actively searched for new particles that would mix with the active neutrinos. We study their phenomenology when they are produced in the evaporation of early Universe black holes, a novel production mechanism that differs from all others and does not depend on the active-sterile mixing. The resulting hotter sterile neutrinos have a distinct spectrum and could be warm dark matter in the 0.3 MeV to 0.3 TeV mass range, distinct from the typical keV range. The possible coincidence of X-rays and gravitational waves is a unique novel signature of our scenario.
\end{abstract}

\maketitle
 
{\it Introduction}.--
Astrophysical observations definitively establish that dark matter (DM) composes about $85\%$ of the matter in the Universe (see e.g.~\cite{Bertone:2016nfn,Gelmini:2015zpa}). Despite decades of efforts to detect their non-gravitational interactions, the nature of the DM remains unknown.  

Sterile neutrinos ($\nu_s$s), proposed fermions that do not transform under the Standard Model (SM) gauge group and only couple to  ``active"  neutrinos $\nu_a$s ($a$ stands for the three flavors $e, \mu$ or $\tau$),   constitute a particularly simple and well-motivated DM candidate, usually if they have $\mathcal{O}$(keV) mass~(see e.g.~\cite{Boyarsky:2018tvu}). For simplicity, we are going to consider a $\nu_s$ with mixing $\sin\theta$ with just one $\nu_a$.

Sterile neutrinos are necessary for see-saw mechanisms~\cite{Yanagida:1979as,Gell-Mann:1979vob,Yanagida:1980xy} that in minimal extensions of the SM  generate the observed~\cite{Super-Kamiokande:1998kpq} small neutrino masses, and they could explain a variety of other intriguing observations.  One of them is the putative 3.5 keV X-ray emission line originating from galaxy clusters~\cite{Bulbul:2014sua,Boyarsky:2014jta} that could be due to $\nu_s \to \nu_a \gamma$ decays of $\nu_s$s with mass $m_s =7.1$~keV and, assuming they constitute all of the DM, mixing  $\sin^2 2\theta \simeq 7 \times 10^{-11}$~\cite{Bulbul:2014sua,Boyarsky:2014jta}, although this interpretation has been challenged~(e.g.~\cite{Jeltema:2014qfa,Dessert:2018qih,Dessert:2023fen}). Heavier decaying $\nu_s$s have been proposed to alleviate the discrepancies of recent Hubble constant measurements~(e.g.~\cite{Gelmini:2019deq,Gelmini:2020ekg,Boyarsky:2021yoh}).

The production of $\nu_s$s in the early Universe through the minimal scenario of non-resonant active-sterile oscillations and collisional processes (so-called ``Dodelson-Widrow'' (DW) mechanism~\cite{Barbieri:1989ti,Barbieri:1990vx,Dodelson:1993je}) is strongly constrained by observations, in particular X-ray data~(e.g.~\cite{Palazzo:2007gz,Horiuchi:2013noa,Perez:2016tcq,Dessert:2018qih,Roach:2022lgo}). Alternative models, such as resonant active-sterile oscillations due to a sizable initial lepton asymmetry, or additional particle interactions (e.g. inflaton decays~\cite{Shaposhnikov:2006xi}, freeze-in in extended particle models~\cite{Asaka:2006ek,Roland:2014vba}, and extended gauge symmetry~\cite{Dror:2020jzy,Bezrukov:2009th}) can result in the production of distinctive cosmological $\nu_s$ abundances and spectra. Predictions are also sensitive to cosmological scenarios~\cite{Gelmini:2019esj,Gelmini:2019wfp,Gelmini:2019clw,Chichiri:2021wvw}.

In this work, we examine for the first time the production through the evaporation of primordial black holes (PBHs) in the early Universe of $\nu_s$s as DM constituents, which we call \textit{PBH sterile neutrinogenesis}.  Black hole evaporation has been studied as a way to produce other DM or dark sector particles mostly or entirely decoupled from the SM, or dark radiation (e.g.~\cite{Fujita:2014hha,Lennon:2017tqq,Morrison:2018xla,Hooper:2019gtx,Baldes:2020nuv,Masina:2020xhk,Gondolo:2020uqv,Sandick:2021gew,Cheek:2021odj,Arbey:2021ysg,Marfatia:2022jiz,Kim:2023ixo}), but not $\nu_s$s. As we demonstrate, PBH sterile neutrinogenesis unlocks novel and distinctive fundamental features and signatures of $\nu_s$s as DM constituents.

There are essential key differences between PBH sterile neutrinogenesis and other $\nu_s$ production mechanisms proposed so far: it is independent  of the active-sterile mixing, since the evaporation is a purely gravitational effect, and $\nu_s$s are produced with a distinctive spectrum and higher energies so that they are hot DM (HDM), or warm DM (WDM) or cold DM (CDM) components (i.e relativistic, becoming or already non-relativistic when galaxy cores start forming, at a temperature $\simeq$ 1 keV)  for larger masses.  Furthermore, this mechanism makes possible a coincidence of X-ray and gravitational wave (GW) signals that would distinguish it from others. 

{\it  Primordial black holes formation and evaporation}--
We consider PBHs within a large range of mass, initial abundance, and evaporation temperature. So we need to ensure that PBHs could be formed with these varied characteristics.   Although PBHs can be formed in the early Universe in many ways~(e.g.~\cite{Zeldovich:1967,Hawking:1971ei,Carr:1974nx,GarciaBellido:1996qt,Green:2000he,Frampton:2010sw,Cotner:2019ykd,Cotner:2018vug,Green:2016xgy,Sasaki:2018dmp,Cotner:2018vug,Cotner:2019ykd,Kusenko:2020pcg,Carr:2020gox,Escriva:2022duf,Lu:2022yuc}) we focus on the collapse of primordial cosmological perturbations at horizon crossing~\cite{Carr:1975qj}, that can lead to all the characteristics we assume. This mechanism can readily appear in a broad range of theories (see e.g.~\cite{Escriva:2022duf} for a recent review), allowing us to discuss PBH sterile neutrinogenesis generally.

PBHs that form through the collapse of horizon scale perturbations in a radiation-dominated epoch at temperature $T_{\rm form}$ have an initial mass (see e.g.~\cite{Fujita:2014hha}) 
\begin{equation}
\label{M-formation}
    M_{\rm PBH} = 1.9\times10^{25}\textrm{g} \left[\frac{\gamma}{0.2}\right] \left[\frac{\textrm{TeV}}{T_{\rm form}}\right]^{2}\left[\frac{106.75}{g_*(T_{\rm form})}\right]^{\frac{1}{2}},
\end{equation}
where $\gamma \leq 1$ is the fraction of the mass within the horizon which ends up in the PBH. While the precise value of $\gamma$ depends on the details of the collapse, as is usual (e.g.~\cite{Fujita:2014hha}) we adopt $\gamma=0.2$~\cite{Carr:1975qj} as our fiducial value.

PBHs evaporate through the emission of Hawking radiation, with a temperature~\cite{Hawking:1974rv}
\begin{equation}
\label{eq:tbh}
    T_{\rm PBH} = \frac{1}{8 \pi M_{\rm PBH}}=1.05\times10^5 \textrm{ GeV} \left[\frac{10^8 {\rm g}}{M_{\rm PBH}}\right]~. 
\end{equation}
For simplicity, we assume a monochromatic PBH mass function throughout.

During evaporation, all particles with mass smaller than $T_{\rm PBH}$ are emitted. The evaporation rate~\cite{MacGibbon:1990zk} increases rapidly as the PBH evaporates, thus most of the emission happens at the end of the PBH lifetime. For simplicity, we make the common assumption of instantaneous reheating, in which the redshift of particles emitted earlier in the evaporation process is negligible, and the average energy of the emitted $\nu_s$s is $\langle p\rangle=6.3 ~ T_{\rm PBH}$~\cite{Baldes:2020nuv}.

We assume that PBHs are formed in a radiation dominated Universe with initial energy density fraction $\beta < 1$. Because they are matter, their energy density fraction increases with time, and PBHs could matter-dominate the Universe before evaporating. In this case, we call ``PD'', the PBH evaporation would reheat the Universe to the temperature $T_{\rm RH}$. PD requires $\beta$ to be large enough for a given PBH mass (see e.g.~\cite{Fujita:2014hha}). If $\beta$ does not reach the value necessary for PD,  the evaporation happens while the Universe is radiation-dominated at temperature $T_{\rm evap}$ (``RD" case). In the PD case, the fraction $f_{\rm evap}$ of the total density in PBHs at evaporation is $f_{\rm evap}=1$ and
\begin{equation}
\label{eq:trhm}
    T_{\rm RH} = 46~{\rm MeV}\left[  \frac{T_{\rm PBH}}{\rm 10^5 GeV} \right]^{\frac{3}{2}}\left[\frac{10.75}{g_*(T_{\rm RH})} \right]^{\frac{1}{4}}\left[\frac{g_H}{110}\right]^{\frac{1}{2}}.
\end{equation}
In the RD case $f_{\rm evap}$ can take any value $f_{\rm evap}<1$ and
\begin{equation}
\label{eq:trhevap}
        T_{\rm evap}=40~{\rm MeV}\left[  \frac{T_{\rm PBH}}{\rm 10^5 GeV} \right]^{\frac{3}{2}}\left[\frac{10.75}{g_*(T_{\rm evap})} \right]^{\frac{1}{4}}\left[\frac{g_H}{110}\right]^{\frac{1}{2}}.
\end{equation}
Here $g_H = 108 + 2$, is the effective number of degrees of freedom (d.o.f.) in the Hawking radiation (of which 108 corresponds to all the SM particles and 2 corresponds to $\nu_s$s)~\cite{Hooper:2019gtx}, and $g_*(T)$  is the density number of d.o.f. at $T$ (coincides with the entropy d.o.f. at the $T$ of interest). 

Rapid reheating in the PD case converts large density fluctuations into radiation, yielding large pressure waves and a quadrupole moment that results in a significant production of  GWs~(e.g.~\cite{Inomata:2019ivs,Inomata:2020lmk,Domenech:2020ssp,Domenech:2021wkk,Domenech:2021ztg}), whose energy density spectrum at the present is approximately~\cite{Domenech:2021ztg, Domenech:2020ssp}
\begin{equation}
\label{eq:gwpresent}
    \Omega_{\rm GW} \simeq \Omega_{\rm GW}^{\rm peak} \left(\frac{f}{f_{\rm UV}}\right)^{5} \Theta(f_{\rm UV}-f)~,
\end{equation}
for  $f<f_{UV}$.  For $f>f_{UV}$ the spectrum decreases very fast and cuts off at $2f_{UV}$. The peak is at $f_{UV}$, 
\begin{align}
\label{fUV}
    f_{\rm UV} =1.7\times 10^{3}\textrm{ Hz}\left[\frac{106.75}{g_*(T_{\rm RH})}\right]^{\frac{1}{12}}
  \left[\frac{10^4~{\rm g}}{M_{\rm PBH}}\right]^{\frac{5}{6}},
\end{align}
the ultraviolet cut-off of the gravitational potential fluctuations power spectrum. The peak GW density is  
\begin{equation}
\label{eq:GWpeak}
    \Omega_{\rm GW}^{\rm peak} = 1.6\times10^{-6} \left[\frac{\gamma}{0.2}\right]^{\frac{7}{9}} \left[\frac{\beta}{10^{-8}}\right]^{\frac{16}{3}}\left[\frac{M_{\rm PBH}}{10^7~{\rm g}}\right]^{\frac{34}{9}}.
\end{equation}

\begin{figure}[t] \centering 
\includegraphics[width=0.35\textwidth]{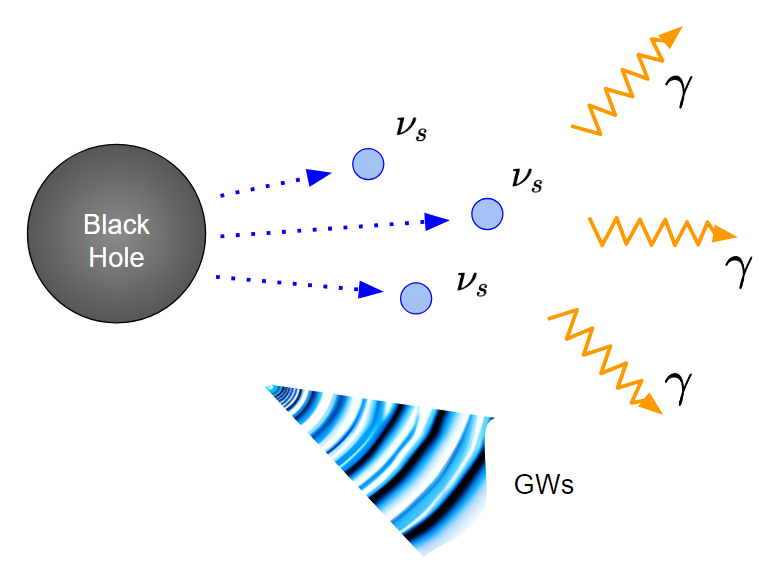}
\caption{Schematic view of PBH sterile neutrinogenesis if PBHs matter-dominate the Universe before evaporating (PD case) that can result in  related X-ray or $\gamma$-ray (from $\nu_s \to \nu_a \gamma$) and gravitational wave (GW) signals.
} 
\label{fig:schematic}
\end{figure}

\begin{figure*}[t] \centering  
\includegraphics[width=0.45\textwidth,valign=c]
{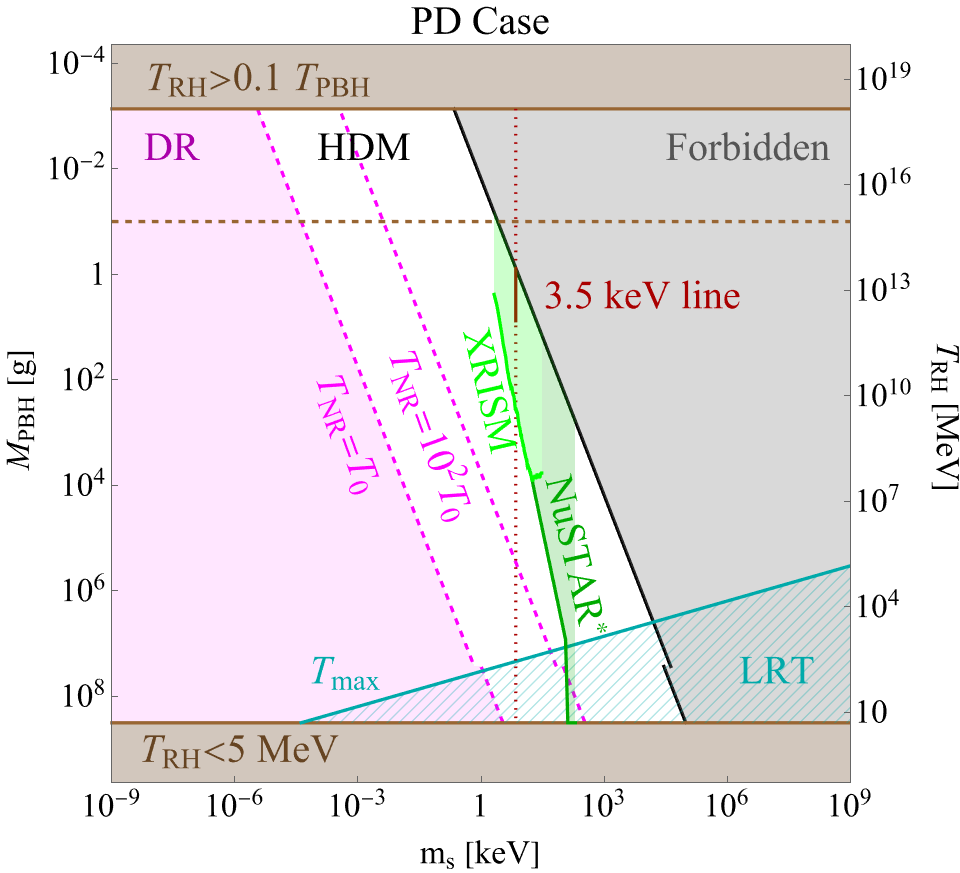}
\begin{minipage}{0.45\linewidth}
\includegraphics[trim={0 0 1.1cm 1cm},clip, width=0.97\textwidth,valign=c]{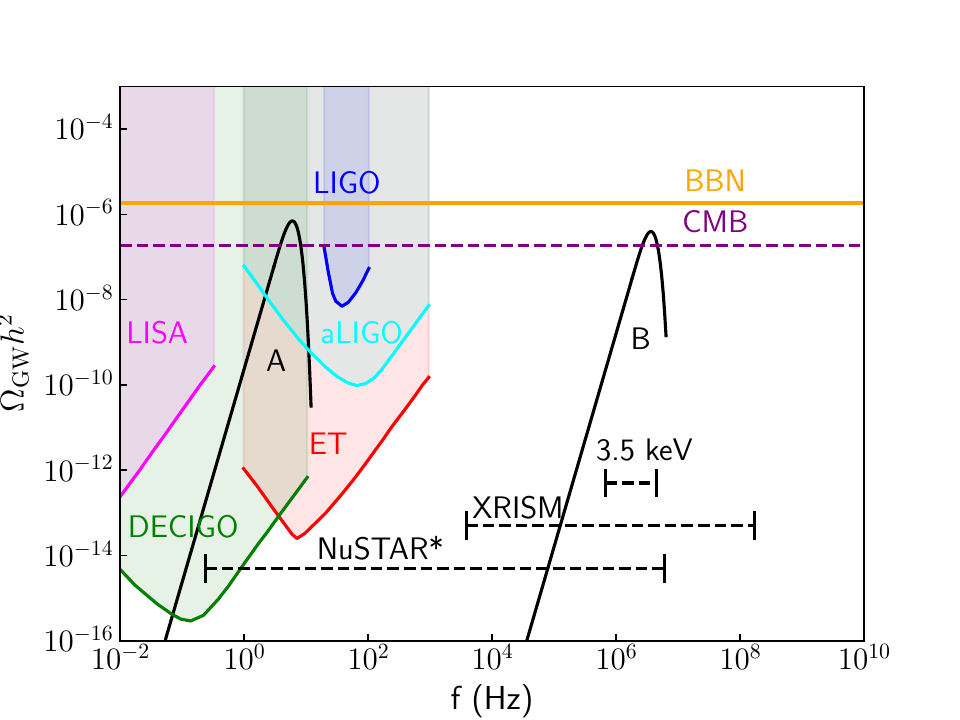}
\end{minipage}
\caption{
\textbf{[Left]} Regions and bounds of interest in the $M_{\rm PBH}$ (left axis) or $T_{\rm RH}$(right axis) vs. $m_s$ space, for $\nu_s$s produced through PBH sterile neutrinogenesis if PBHs matter-dominate the Universe before evaporating (which we call PD), where $\nu_s$s would be dark radiation (light magenta,``DR''), HDM (white) or forbidden (grey) by the Lyman-$\alpha$ forest limit. Brown regions are excluded by requirements on $T_{\rm RH}$. When $T_{\rm RH} <T_{\rm max}$ this a
low reheating temperature scenario (hatched in cyan, ``LRT''). In the green regions $\nu_s \to \nu_a \gamma$ decays of $\nu_s$s dominantly produced by PBH evaporation could lead to observable X-ray lines, the putative 3.5 keV (brown line) and detectable in the future by an improved NuSTAR-like experiment~\cite{Ng:2019gch}) (darker green, NuSTAR*) or XRISM~\cite{XRISMScienceTeam:2020rvx} (lighter green). Two values of $T_{\rm NR}$, the temperature when $\nu_s$s become non-relativistic, are indicated (dotted magenta lines). 
The region above the horizontal dashed brown line can be restricted by the upper bound on the energy scale of inflation $\rho_{\rm inf} < (1.6 \times 10^{16} \textrm{ GeV})^4$~\cite{Planck:2018jri}.
\textbf{[Right]} Gravitational wave (GW) density as function of frequency induced by PBH instantaneous reheating in the PD scenario. Target peak frequency ranges of possible coincident GW and X-ray signatures are indicated (dashed black lines), together with two examples of possible GW signals in Eq.~\eqref{eq:gwpresent} (solid black lines), for models  A ($M_{\rm PBH}=2\times10^7$~g, $\beta=6\times10^{-9}$) and B ($M_{\rm PBH} = 1$~g, $\beta =8\times10^{-4}$) as well as sensitivity curves for LISA, DECIGO, ET and Advanced LIGO searches~\cite{Thrane:2013oya}, and 95\% C.L. limits on $\Delta N_{\rm eff}$ from BBN~\cite{Arbey:2021ysg}, and projections from  CMB-S4~\cite{Abazajian:2019eic}. 
} 
\label{fig:parameterspace}
\end{figure*}

The potential combination of an X-ray line observation in galaxies and galaxy clusters due to two body $\nu_s$ decays and a GW signal would be a unique signature of PBH sterile neutrinogenesis in the PD case, as shown schematically in Fig.~\ref{fig:schematic} and in detail in Fig.~\ref{fig:parameterspace}.

\begin{figure}
    \centering
\includegraphics[width=0.45\textwidth]{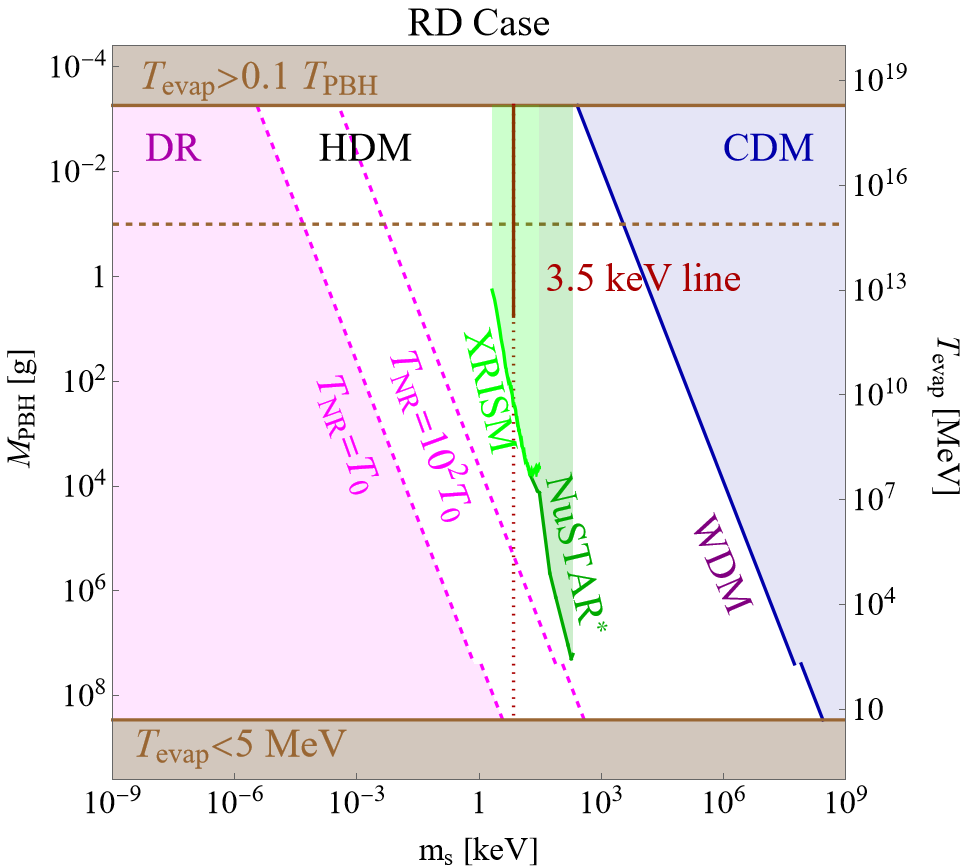}
    \caption{Regions and bounds of interest the $M_{\rm PBH}$ (left axis) or $T_{\rm evap}$ (right axis) vs. $m_s$ space, for $\nu_s$s produced through PBH sterile neutrinogenesis if PBHs evaporate during radiation domination (RD case), where $\nu_s$s would be dark radiation (light magenta ``DR''), HDM (white), WDM (blue line along the Lyman-$\alpha$ forest limit), or CDM (blue). In the last two, $\nu_s$s can account for all of the DM. Brown regions are excluded by requirements on $T_{\rm evap}$.  In the green regions $\nu_s \to \nu_a \gamma$ decays of $\nu_s$s dominantly produced by PBH evaporation could lead to observable X-ray lines, the putative 3.5 keV (brown line) or detectable in the future by an improved NuSTAR-like experiment~\cite{Ng:2019gch}) (darker green, NuSTAR*) or  XRISM~\cite{XRISMScienceTeam:2020rvx} (lighter green).  Two values of $T_{\rm NR}$, the temperature at which $\nu_s$s become non-relativistic, are indicated (dotted magenta lines). The region above the horizontal dashed brown line can be restricted by the upper bound on the energy scale of inflation $\rho_{\rm inf} < (1.6 \times 10^{16} \textrm{ GeV})^4$~\cite{Planck:2018jri}.
    }
 \label{fig:Raddom} 
\end{figure}

We require that $T_{\rm RH}$ or $T_{\rm evap}$ be larger than 5 MeV due to limits imposed by Big-Bang Nucleosynthesis (BBN)~\cite{Kawasaki:1999na,Kawasaki:2000en,Hannestad:2004px,Ichikawa:2005vw,Ichikawa:2006vm,DeBernardis:2008zz,deSalas:2015glj,Hasegawa:2019jsa}, and be smaller than 0.1~$T_{\rm PBH}$, since for the consistency of our model they cannot be as large  the PBH temperature at evaporation. These conditions imply that $T_{\rm RH} <1.4\times 10^{15}$ GeV and $7.3\times 10^{-4}$g$<M_{\rm PBH} <4.7\times 10^{8}$g in the PD case, and $T_{\rm evap} <1.9\times 10^{15}$GeV, and  $5.4\times 10^{-4}$g $<M_{\rm PBH}<4.2\times 10^{8}$g in the RD case. This is the range of parameters in the vertical axis of Fig.~\ref{fig:parameterspace} (left panel) and  Fig.~\ref{fig:Raddom} for PD and RD respectively. The region above the dashed horizontal line in these figures
can be restricted by 
the upper bound on the energy scale of inflation $\rho_{\rm inf} < (1.6 \times 10^{16} \textrm{ GeV})^4$~\cite{Planck:2018jri}.

\textit{PBH sterile neutrinogenesis.}---
During PBH evaporation, $\nu_s$s are produced if $m_s < T_{\rm PBH}$. Since we impose $M_{\rm PBH} < 4\times 10^8$g, i.e. $T_{\rm PBH}>10^4$GeV,  $\nu_s$s with mass $m_s <10^4$ GeV are emitted ultra-relativistic. The density of $\nu_s$s relative to the total energy density  at emission, considering the fraction of the Hawking radiation going into $\nu_s$, $2/ g_H=1/55$, is
\begin{equation}
\label{eq:nus-abundance-at-evap}
\left.\frac{\rho_s}{\rho_{\rm total}}\right|_{\rm evap} =  \frac{f_{\rm evap}}{55}~.
\end{equation}

We find that except when the active-sterile mixing is  $\sin^2 2\theta \geq 5\times 10^{-6}({\textrm{keV}/{m_s}})$, 
$\nu_s$s do not thermalize and just redshift after they are produced. Requiring that PBH sterile neutrinogenesis is dominant with respect to the DW production (Eqs.~\eqref{eq:sinMstd} and \eqref{eq:sinMLRT}) ensures this regime, so the present fraction of the DM in $\nu_s$s is
\begin{equation}
\label{eq:fnr-MD}
     f_s \simeq 2\times10^{-6} f_{\rm evap} \left[\frac{m_s}{\textrm{ keV}}\right]\left[\frac{10^8{\rm g}}{M_{\rm PBH}}\right]^{\frac{1}{2}}\left[\frac{10.75}{g_*(T_{\rm evap})}\right]^{\frac{1}{4}}
\end{equation}
in the RD case, and for PD we take $f_{\rm evap}=1$ and replace $T_{\rm evap}$ by $T_{\rm RH}$. Moreover, the  $\nu_s$ spectrum redshifts keeping the shape it has at production (shown e.g. in Fig.~1 of \cite{Baldes:2020nuv}), very different than in other production models.

PBH evaporation is the dominant production mechanism only if
the DW DM density fraction $f_{s,osc} \propto \sin^2 2\theta$ 
(see e.g.~\cite{Gelmini:2019wfp}) is $f_{s,osc} < f_{s}$, which for non-relativistic sterile neutrinos requires
\begin{equation}     
\label{eq:sinMstd}
 \frac{\sin^2 2\theta}{
    4\times 10^{-13}} < f_{\rm evap} \frac{{\rm keV}}{m_s} \left[\frac{10.75}{g_*(T_{\rm evap})}\right]^{\frac{1}{6}}\left[\frac{g_*(T_{\rm max})}{30}\right]^{\frac{3}{2}}\left[\frac{T_{\rm evap}}{\rm 5 MeV}\right]^{\frac{1}{3}}.
\end{equation}
Here $T _{\rm max}$  is the temperature at which the DW production rate has a sharp maximum (see e.g.~\cite{Dodelson:1993je,Gelmini:2019wfp}),
\begin{equation}     
\label{eq:Tmax}
  T _{\rm max}\simeq 133~\textrm{MeV}\left[\frac{m_s}{\rm keV}\right]^{1/3},
\end{equation}
 and in the PD case $T_{\rm evap}$ is replaced by $T_{\rm RH}$. 

When $T_{\rm RH} < T_{\rm max}$ the PD scenario is a realization of the low reheating temperature (LRT) model~\cite{Gelmini:2004ah,Gelmini:2019esj,Gelmini:2019wfp} (see region  hatched in cyan in Fig~\ref{fig:parameterspace}), in which the DW production is suppressed and the condition for 
 PBH sterile neutrinogenesis to be dominant is 
\begin{equation}
\label{eq:sinMLRT}
\frac{\sin^2 2\theta}{10^{-9}}< f_{\rm evap}\left[\frac{10.75}{g_*(T_{\rm RH})}\right]^{\frac{1}{6}}\left[ \frac{\rm 5~{\rm MeV}}{T_{\rm RH}}\right]^{\frac{8}{3}}.
\end{equation}

For sterile neutrinos heavier than the active neutrinos $m_s>m_a$, the high average momentum, $\langle p \rangle= \langle \epsilon \rangle T_{\rm evap}\simeq 6.3~ T_{\rm PBH}$~\cite{Baldes:2020nuv}, with which $\nu_s$'s are produced in the PBH sterile neutrinogenesis process can result in a relativistic population at present if $m_s< \langle \epsilon \rangle T_0 [3.90/g_*(T_{\rm evap})]^{1/3}$, where $T_0 = 2.3\times10^{-4}\textrm{ eV}$ is the present radiation temperature. In this case, the condition $f_{s,osc} < f_{s}$  for sterile neutrinogenesis to be dominant is
\begin{equation}
    \frac{\sin^2 2\theta}{2.9\times10^{-7}} < f_{\rm evap}\left[\frac{\textrm{eV}}{m_s}\right]^2\left[\frac{10.75}{g_*(T_{\rm evap})}\right]^{\frac{1}{3}}\left[\frac{g_*(T_{\rm max})}{10.75}\right]^{\frac{3}{2}}.
\end{equation}

If, however, sterile neutrinos are lighter than active neutrinos, $m_s< m_a$, then the negative mass difference $\delta m^2 = m_a^2 - m_s^2$ can induce resonant production of sterile neutrinos leading to a sizeable lepton asymmetry in the active neutrino just before BBN~\cite{Dolgov:2002wy}. For the small mixing angles relevant to our study, the generated asymmetry is approximately (see below Eq.~(383) in Ref.~\cite{Dolgov:2002wy})
\begin{equation}
    L_a\simeq 30 \left[\frac{\delta m^2}{2.5\times10^{-3}\textrm{ eV}^2}\right]\left[\frac{\sin^2 2\theta}{10^{-8}}\right]^4.
\end{equation}
Considering the BBN bounds on the lepton asymmetry, $L_e < 2\times10^{-3}, L_{\mu,\tau}<0.1$~\cite{Barenboim:2016shh},  the most stringent case of sterile-electron neutrino mixing, implies for light sterile neutrinos $\sin^2 2\theta \lesssim 9\times10^{-10} (\delta m^2/2.5\times10^{-3}\textrm{ eV}^2)$.

The lepton asymmetry can be related to the number density of sterile neutrinos produced during this resonance phase by $n_{\nu_s} \simeq L_{a}n_\gamma$. The sterile neutrinos produced resonantly will typically have a cold spectrum with low average momentum, which we approximate with $\langle \epsilon \rangle \simeq 1$, in contrast to the hot spectrum produced by sterile neutrinogenesis. If at present, $T_{\nu}< m_s$, the resonantly-produced sterile neutrinos are non-relativistic and the condition for PBH sterile neutrinogenesis to be dominant is
\begin{equation}
    \frac{\sin^2 2\theta}{6.4\times10^{-10}} <  f_{\rm evap}^{\frac{1}{4}} \left[\frac{0.01\textrm{ eV}}{m_s}\right]^{\frac{1}{4}} \left[\frac{2.5\times10^{-3}\textrm{ eV}^2}{\delta m^2}\right]^{\frac{1}{4}},
\end{equation}
and if at present $T_{\nu}>m_s$, then the resonantly-produced sterile neutrinos are relativistic and the condition becomes
\begin{equation}
    \frac{\sin^2 2\theta}{1.8\times10^{-9}} <  f_{\rm evap}^{\frac{1}{4}} \left[\frac{2.5\times10^{-3}\textrm{ eV}^2}{\delta m^2}\right]^{\frac{1}{4}}.
\end{equation}

\textit{Constraints and possible signatures}--
In the PD case, we find that $\nu_s$s can only be a HDM component (as was found for other DM candidates, unless a large late entropy production is assumed, see e.g.~\cite{Fujita:2014hha}),  thus we must impose $f_s <0.1$. In the RD case, we find instead that they can be WDM or CDM with $f_s=1$, namely, they can account for all of the DM, besides HDM.   This is determined by the Lyman-$\alpha$ forest limit on WDM. High-velocity DM particles are constrained by observations because their free streaming impedes the formation of structure in the Universe.   The $2\sigma$ Ly-$\alpha$ limit~\cite{Baur:2017stq, Villasenor:2022aiy}  excludes thermal fermions (i.e. fermions with a thermal equilibrium distribution) lighter than $m_{\rm therm} = 3$ keV from constituting all of the DM. Close to this limit, thermal particles would be WDM, and sufficiently above it would be CDM.

We translate this limit on thermal fermions, whose average momentum is 3.15~$T$, to $\nu_s$s produced in PBH evaporation whose average momentum at production is $\langle p \rangle= \langle \epsilon \rangle T_{\rm evap}= 6.3 T_{\rm PBH}$~\cite{Baldes:2020nuv}.
We obtain~\cite{Baur:2017stq,Gelmini:2019wfp}
\begin{align}
\label{eq:massrelation}
    m_s \geq 4.1\textrm{ keV}\left[\frac{\langle \epsilon\rangle}{3.15}\right] \left[\frac{m_{\rm therm}}{\textrm{ keV}}\right]^{\frac{4}{3}} \left[\frac{10.75}{g_\ast(T_{\rm evap})}\right]^{\frac{1}{3}}, 
\end{align}
  which using Eq.~\eqref{eq:tbh} and Eq.~\eqref{eq:trhm} or  Eq.\eqref{eq:trhevap} (for the PD case $T_{\rm evap}$ should be replaced by $T_{\rm RH}$) becomes
\begin{equation}
\label{eq:lymanR}
    \frac{m_s}{\rm keV}\geq 2.7 \times 10^8\left[ \frac{5 \,\rm MeV}{T_{\rm evap}} \right]^{\frac{1}{3}}\left[\frac{10.75}{g_\ast(T_{\rm evap})}\right]^{\frac{1}{6}}. 
\end{equation}
We checked that this limit coincides with that~\cite{Bode:2000gq, Fujita:2014hha, Baldes:2020nuv,Decant:2021mhj} derived from the upper bound on the present average speed of WDM particles.

In the PD case, using Eq.~\eqref{eq:lymanR} in Eq.~\eqref{eq:fnr-MD} with $f_{\rm evap}=1$ we find that $\nu_s$s with mass above the Ly-alpha limit would necessarily have $f_s >1$ and are thus forbidden. Below this limit, $\nu_s$s from PBH neutrinogenesis can only be an HDM component, thus with  $f_s<0.1$, or dark radiation (DR) if they are still relativistic at present. The corresponding regions in the $M_{\rm PBH}$ or $T_{\rm RH}$ versus $m_s$ space are shown in grey, white, and light magenta respectively in the left panel of Fig.~\ref{fig:parameterspace}.

In the RD case, using  Eq.~\eqref{eq:lymanR} in Eq.~\eqref{eq:fnr-MD} with $f_{\rm evap}<1$, $\nu_s$s with mass above the Ly-alpha limit can have  $f_s =1$, and be WDM very close to the limit  or CDM above it, as shown in Fig.~\ref{fig:Raddom} (blue boundary line and region respectively). As seen in the figure, $\nu_s$s can be  WDM in the 0.3 MeV to 0.3 TeV mass range, and account for all of the DM, $f_s=1$, if $f_{\rm evap}=4.04\times 10^{-3}[g_*(T_{\rm evap})/10.75]^{1/3}$. Their mixing needs to be below an upper limit that goes between $10^{-24}$ for $m_s \simeq$ 0.3 TeV to $10^{-11}$ close to 0.3 MeV for DW production to be subdominant and should be as small as needed to escape all observational limits (such as those on ``heavy neutral leptons with a neutrino portal to the SM"~\cite{Abdullahi:2022jlv} as $\nu_s$s are sometimes called). 

In our model $\nu_s$s  become non-relativistic at a temperature $T_{\rm NR}$, two values of which  are indicated in Fig.~\ref{fig:parameterspace} (left panel) and  Fig.~\ref{fig:Raddom}. When $T_{\rm NR}< T_0$, $\nu_s$s are dark radiation. They are  WDM  when $T_{\rm NR} \simeq 1$ keV, which  approximately coincides with the Ly-$\alpha$ limit.  

 The $\nu_s$ decay rate is proportional to $\sin^2\theta$. Those $\nu_s$s lighter than the $\pi_0$, with $m_s < 135$~MeV, decay  mostly into three active neutrinos, $\nu_s \rightarrow 3\nu_a$  (e.g.~\cite{Barger:1995ty, Fuller:2011qy}), but have also a decay mode $\nu_s \to \nu_a \gamma$ with branching ratio $\simeq 1\%$ into photons of energy $m_s/2$, and this mode could lead to prominent X-ray and $\gamma$-ray line signals in astrophysical searches. Such a signal would correspond to a measured photon flux proportional to $f_s~ \Gamma_{\nu_s \to \nu_a \gamma}\sim f_s \sin^2 \theta$, and since in PBH sterile neutrinogenesis the fraction $f_s$ of the DM in $\nu_s$s is independent of the mixing, the parameter space corresponding to possible observations is different from that in other production scenarios in which $f_s$ depends on the mixing.

 Fig.~\ref{fig:parameterspace} (left panel) and  Fig.~\ref{fig:Raddom} show in green  the parameter region where $\nu_s$s in our model could account for an observed  X-ray line. In the lighter green region a $\nu_s$ signal detectable by the upcoming X-Ray Imaging and Spectroscopy Mission
(XRISM) satellite~\cite{XRISMScienceTeam:2020rvx} (sensitive to X-ray energies $\sim 0.4-15$~keV) could be due to PBH sterile neutrinogenesis.  In the darker green the same would be possible for a future NuSTAR-like experiment we call ``NuSTAR*'', with twice the sensitivity of NuSTAR~\cite{Ng:2019gch} in the $\sim 10-200$~keV band. A detected $\nu_s$ decay in these regions would have a noticeably different abundance-mixing angle relation than that predicted in other production mechanisms. The  
 brown line corresponds to the putative $3.5$ keV signal.
 The green region is bounded from above either by the Lyman-$\alpha$ limit (in the PD case) or the upper limit on $T_{\rm evap}$ (in the RD case), and from below by requiring that PBH neutrinogenesis dominates over DW production. In these regions $f_s> 10^{-4}$, thus the $\nu_s$ lifetime is always much larger than the age of the Universe,  $\tau \gg t_U$.    Notice that in this region $\nu_s$s are already non-relativistic at $T_{\rm NR} > 10^2~ T_0$.
 
If $\nu_s$s produced in PBH sterile neutrinogenesis are still relativistic when the microwave background (CMB) is emitted, i.e. $T_{\rm NR} < 10^3~ T_0$, their density relative to the total would be the same as in Eq.~\eqref{eq:nus-abundance-at-evap}. Thus, they would contribute to the effective number of neutrino species $N_{\rm eff}$ at that epoch by $\Delta N_{\rm eff} \simeq 6.8\times10^{-2} f_{\rm evap}$ (e.g.~\cite{Hooper:2019gtx}). Even in the PD case, where $f_{\rm evap}=1$, this  contribution is below current bounds~\cite{Planck:2018vyg},  but could be probed by CMB-S4, whose expected sensitivity is $\Delta N_{\rm eff} \leq 0.06$~\cite{Abazajian:2019eic}. Moreover, in the PD case GWs from PBH evaporation would also contribute to  $\Delta N_{\rm eff}$.

We explore now the possible coincidence of  X-ray  and GW signals.
The $M_{\rm PBH}$ range of the X-ray regions (green regions and brown line) in the left panel of Fig.~\ref{fig:parameterspace} directly translates through Eq.~\eqref{fUV} into a GW peak frequency range shown in the right panel of the same figure with dashed lines, for the putative 3.5 keV line and possible lines that may be observed by XRISM~\cite{XRISMScienceTeam:2020rvx} or NuSTAR* (an imagined upgrade of NuSTAR~\cite{Ng:2019gch}).   The right panel of Fig.~\ref{fig:parameterspace} also shows present limits and future reach of GW observatories. Two examples of GW spectrum in Eq.~\eqref{eq:gwpresent} with values of $\beta$ in Eq.~\eqref{eq:GWpeak} were chosen to produce GW peak amplitudes below the BBN limit on the GWs contribution  to $\Delta N_{\rm eff}$~\cite{Arbey:2021ysg} (see Fig. 4 of Ref.~\cite{Domenech:2020ssp}), and above the expected reach of CMB-S4~\cite{Abazajian:2019eic}. Model A has $M_{\rm PBH}=2\times 10^7$g and $\beta=6\times10^{-9}$, and model B, $M_{\rm PBH} = 1$~g and $\beta =8\times 10^{-4}$.  A NuSTAR* experiment could find X-ray signals correlated with a GW signal in the frequency range observable by advanced LIGO, the Einstein Telescope and DECIGO~\cite{Thrane:2013oya}, as in model A.  The 3.5 keV X-ray emission line and a possible signal in XRISM could correlate with a GW background contributing a  $\Delta N_{\rm eff}$ measurable by CMB-S4,  if the initial PBH abundance $\beta$ is large enough, as in example B.  

\textit{Conclusions.}--- Sterile neutrinos that mix with SM active neutrinos are particularly well motivated beyond SM physics particles. We point out that if they are produced in the evaporation of PBHs their astrophysical and cosmological characteristics could be very different than if they are produced through all other mechanisms studied so far. PBH sterile neutrinogenesis, which can be implemented in a variety of cosmological theories, depends only on the $\nu_s$ mass. Thus, $\nu_s$s can contribute a sizable fraction to, or all of, the  DM for arbitrarily small active-sterile mixing angles, and $\nu_s$s have a distinct spectrum much hotter at production. 

 Sterile neutrinos in the keV range are usually considered good WDM candidates, but  are HDM in our scenario and could be WDM in the 0.3 MeV to 0.3 TeV mass range instead.
 
 If PBHs matter-dominated the Universe before evaporating, instantaneous reheating yields GWs with a characteristic frequency spectrum that could correlate with an X-ray or $\gamma$-ray signal due to $\nu_s$ decays, either in upcoming instruments or if the 3.5 keV possible signal is confirmed. The presence of both signals would be a unique signature of our scenario.  

 Our work opens novel avenues to explore the fundamental properties of $\nu_s$s and connect them with other fields of research.

~\newline
\textit{Acknowledgements. - } G.G. and M.C. were supported in part by the U.S. Department of
Energy (DOE) Grant No. DE-SC0009937 and M.C. also by the UC Southern California Hub, funded by the UC National Laboratories division of the University of California Office of the President. P.L. was supported by Grant Korea NRF2019R1C1C1010050.
V.T. acknowledges support by the World Premier International Research Center Initiative (WPI), MEXT, Japan and JSPS KAKENHI grant No. 23K13109.

 
\bibliography{references}
 
\end{document}